\documentclass[9pt,conference]{IEEEtran}

\usepackage{waspaa25} 

\usepackage{bm} 

\usepackage[utf8]{inputenc} 
\usepackage[T1]{fontenc}    
\usepackage{hyperref}       
\usepackage{url}            
\usepackage{booktabs}       
\usepackage{amsfonts}       
\usepackage{nicefrac}       
\usepackage{microtype}      
\usepackage{xcolor}         

\usepackage{stfloats}       
\usepackage{graphicx,times}
\usepackage{etoolbox,siunitx}
\robustify\bfseries
\usepackage{balance}
\usepackage{amssymb}
\usepackage{bm}
\usepackage{tabularx}
\usepackage{caption}
\usepackage{multirow}
\usepackage{mathtools}
\usepackage{marginnote}
\usepackage{cite}
\usepackage{pifont}
\usepackage[normalem]{ulem}

\usepackage[hang,flushmargin]{footmisc}

\definecolor{nicegreen}{rgb}{0.1, 0.6, 0.2}

\newcommand{\paramT}[1]{$\text{T}_{#1} $}  
\newcommand{\paramC}[1]{$\text{C}_{#1} $}  
\newcommand{\paramDRR}[0]{$\text{DRR} $}  %
\newcommand{\paramEDT}[0]{$\text{EDT} $}  %

\def\taskname{Scene-wide Acoustic Parameter Estimation}

\title{\taskname}

\name{Ricardo Falcon-Perez$^{1}$, 
      Ruohan Gao$^{2}$,
      Gregor Mueckl$^{3}$,
      Sebastia V. Amengual Gari$^{3}$,
      Ishwarya Ananthabhotla$^{3}$}
\address{$^{1}$Aalto University, Finland \;
$^{2}$University of Maryland, USA \;  
$^{3}$Meta Reality Labs Research, USA
}

\begin{document}

\maketitle

\begin{abstract}
For augmented (AR) and virtual reality (VR) applications, accurate estimates of the acoustic characteristics of a scene are critical for creating a sense of immersion.  However, directly estimating Room-impulse Responses (RIRs) from scene geometry is often a challenging, data-expensive task.  We propose a method to instead infer spatially-distributed acoustic parameters (such as C50, T60, etc) for an entire scene from lightweight information readily available in an AR/VR context.  We consider an image-to-image translation task to transform a 2D floormap, conditioned on a calibration RIR measurement, into 2D heatmaps of acoustic parameters.  Moreover, we show that the method also works for directionally-dependent  (i.e. beamformed) parameter prediction.  We introduce and release a 1000-room, complex-scene dataset to study the task, and demonstrate improvements over strong statistical baselines. 
\end{abstract}












\section{Introduction}

As sound propagates in indoor spaces, it bounces and interacts with surfaces such as room boundaries, furniture, and other objects, creating unique acoustic impressions \cite{kuttruff2009}. For example, long reverberation times can enhance the enjoyment of chamber music, while also reducing the clarity of speech; and strong directional reflections can degrade source localization ability \cite{nabelek1989,blauert1996}. Estimating the acoustic properties of scenes can be useful for various applications, such as architectural planning \cite{architectural_acoustics}, acoustic design or treatment of spaces for specific activities \cite{Laukkanen2014}, and augmented reality (AR) and virtual reality (VR) \cite{yang2022audio}.

In AR/VR, accurate estimates of acoustics are essential for perceptually plausible rendering, which in turn enhances immersion \cite{potter2022, larsson2005, Khairul_Anuar_2024, Vorländer2020}. However, in many virtual environments, we often lack the comprehensive scene information needed for inferring acoustics, such as material properties. Moreover, we need to estimate the acoustic properties of the entire scene so that both sources and receivers can move freely. While accurate acoustics can be obtained by measuring room impulse responses (RIRs), or blindly estimated from signals such as reverberant speech, this is typically limited to a single pair of source-receiver positions, from which acoustics are extrapolated to a whole scene. In complex scenes, like multi-room apartments, extrapolation may be inaccurate, and multiple measurements are required.

RIRs can be also be estimated directly. Physical simulation such as geometric acoustics or wave-based methods can be very accurate, but are computationally expensive, require detailed scene data, and need to be applied to each scene \cite{savioja2015}. Recently, learning-based approaches have used multimodal information that describes the geometry and material properties of the scene \cite{ChenGCG22, mesh2ir, MajumderCAG22, LuoDT00G22, SuCS22,liu2025haae} to infer RIRs. While promising, previous work is limited by poor generalization to new scenes, poor handling of complex, real-world geometries, and single-channel RIR estimation that ignores directional dependencies.

\begin{figure}[htb]
      \centering
      \includegraphics[trim={0cm 0cm 0.5cm 1.0cm},clip, width=0.78\columnwidth]{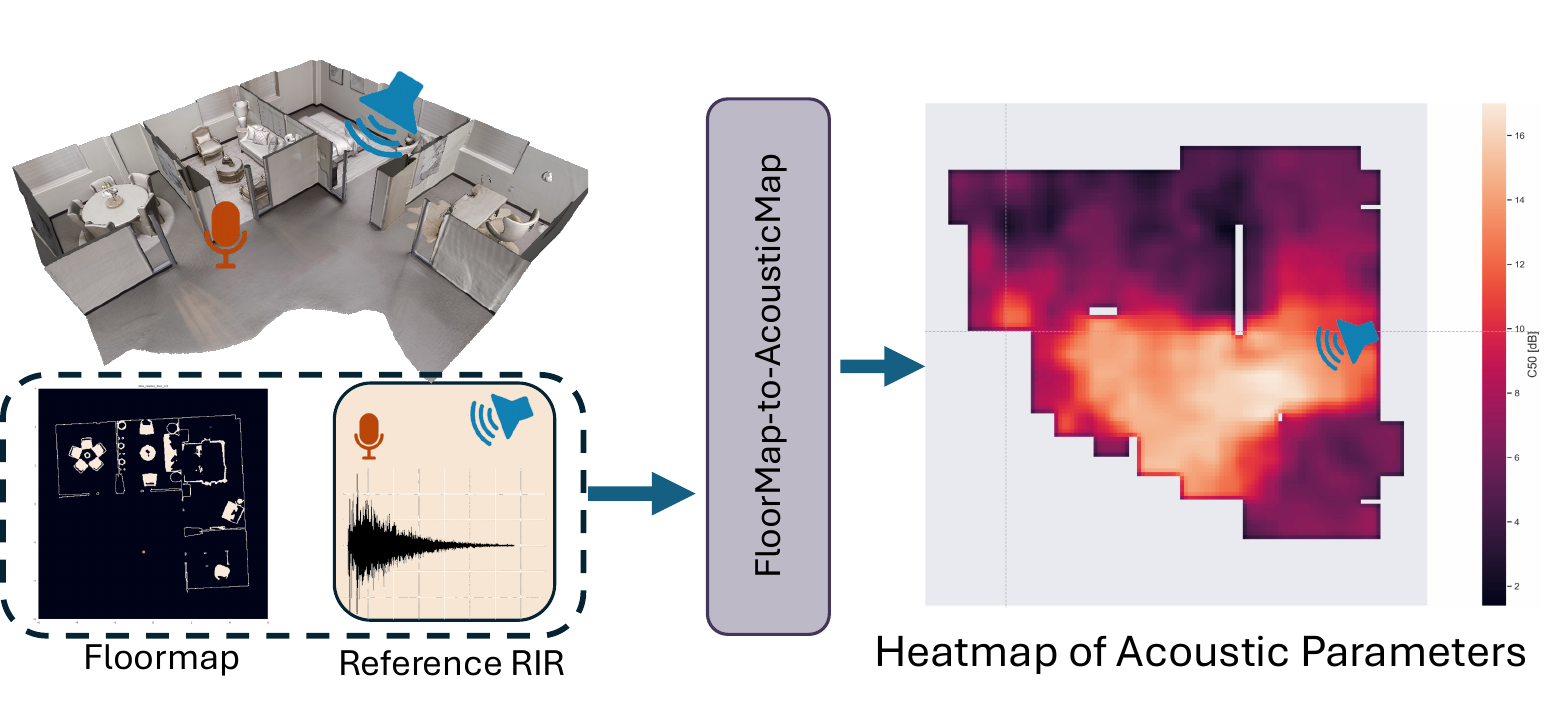}
    \captionsetup{width=0.98\linewidth}
    \caption{We estimate 2D spatially-distributed acoustic parameters for an unseen scene as an image-to-image translation task, using a floormap and a reference RIR as geometric and acoustic inputs.}

    \label{fig:concept}
\end{figure}

An alternative approach is to predict acoustic parameters. Acoustic parameters are standardized metrics that measure specific properties of RIRs, and describe key characteristics of indoor environments, such as reverberation time or speech intelligibility. While these parameters can be computed directly from RIRs, an interesting research area focuses on estimating them from other inputs when RIRs are not available. Reverberation time, for instance, can be roughly predicted from room geometry and absorption coefficients using simple formulas \cite{sabine1922collected, neubauer2001estimation, eyring1930reverberation}, or from spherical maps of absorption with a machine learning model \cite{falcon2019, falcon2021}. Additionally, acoustic parameters can also be estimated from reverberant audio, usually speech, in a process known as blind estimation. Typical models for this task include transformers \cite{Wang2024, wang2024ssbrpeselfsupervisedblindroom}, CRNNs \cite{LopezCC21}, CNNs \cite{icassp/IckMJ23}, and variational autoencoders \cite{gotz2023, götz2024blindacousticparameterestimation}. Finally, some studies explore related tasks that estimate geometrical properties of the room \cite{wang2024berp}, including room dimensions \cite{yuaxin2023}, materials \cite{yu2021, Dilungana2022}, or identify rooms \cite{peters2012}. Previous work focuses on estimating acoustic parameters for a single pair of source and receiver locations; in contrast, our work aims to estimate the acoustic parameters scene-wide in a single inference step, for an unseen scene and arbitrary source position.

In this paper, we propose a method to predict spatially-distributed acoustic parameters for a whole, unseen scene, using limited information easily available in an AR/VR context, as shown in \cref{fig:concept}. We use a 2D floormap of the scene as basic geometric information (e.g. without any material properties), as well as a single, randomly chosen RIR as a calibration input to ground the model's understanding of the acoustic environment. We frame this task as an image-to-image translation problem, where we translate 2D floormaps into 2D heatmaps of acoustic parameters. To study this, we present and release a new, large-scale dataset, called MRAS (Multi-Room Apartment Simulations).  Finally, we demonstrate that our model outperforms algorithmic baselines, and extend our model to showcase spatially-dependent (beamformed) acoustic parameter prediction. \footnote{Code, dataset: \url{https://github.com/facebookresearch/SceneAcousticEstimation}}

\section{Method}
\label{sec:method}

\begin{figure*}[htb]
      \centering
    \includegraphics[trim={0cm 0.0cm 0cm 0.0cm},clip, width=0.70\linewidth]{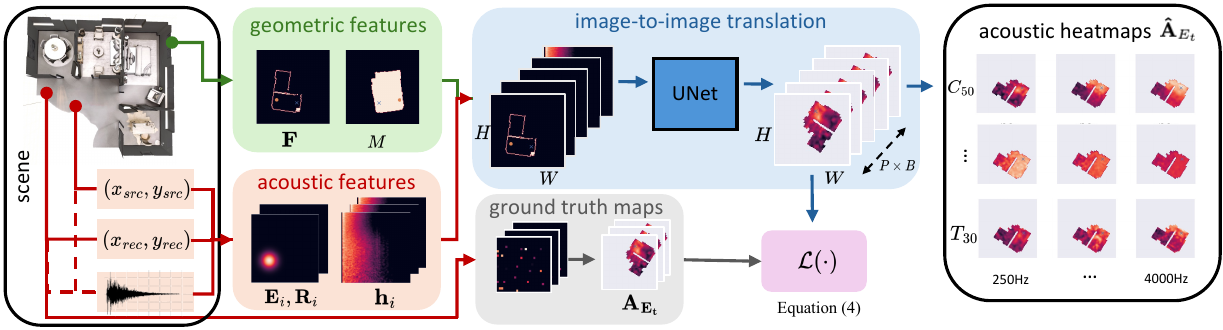}  
    \captionsetup{width=0.99\linewidth}
        \caption{\textbf{Estimation of spatially-distributed acoustic parameters as an image translation task}. Geometric features include 2D floormaps and a mask delimiting the scene area (extracted from the scene mesh or elsewhere). These contain no information about materials or acoustics. An RIR from an arbitrary source-receiver position pair provides acoustic context for the scene, where the source position is also the source for the target heatmap. These are fed to a neural network to predict acoustic heatmaps for $P$ parameters at $B$ frequency bands.}
    \label{fig:system}
\end{figure*}

\subsection{Task Definition}

Our goal is to predict an acoustic parameter heatmap for the entire scene, corresponding to any arbitrary source position, given some reference information about the scene's geometry and acoustics. For a given 3D sound scene $\mathcal{D}$, we denote a sound source (or \textit{emitter}) location as $\mathbf{E}_i \in \mathbb{R}^2$, a receiver location as $\mathbf{R}_i \in \mathbb{R}^2$, and the corresponding room impulse response as a function $ \mathcal{I} : \mathbb{R}^2 \times \mathbb{R}^2 \rightarrow \mathbb{R}^{N \times T}$,
which maps a pair of emitter and receiver locations to an impulse response $\mathbf{h}_i = \mathcal{I}(\mathbf{E}_i, \mathbf{R}_i)$, of $N$ channels and $T$ time steps. Formally, we aim to learn a function $\Phi$ that maps scene and acoustic context to a predicted acoustic map:
\begin{align}
\label{eq:task2}
& \mathbf{h}_r = \mathcal{I}(\mathbf{E}_r, \mathbf{R}_r), \\
& \Phi \colon (\mathbf{F}_D, \mathbf{E}_r, \mathbf{R}_r, \mathbf{h}_r, \mathbf{E}_t) \mapsto \hat{\mathbf{A}}_{E_t},
\end{align}
where the subscripts $r$ and $t$ denote reference and target locations; $\mathbf{F}_D \in \{0,1\}^{H \times W}$ is a binary 2D floormap with height $H$ and width $W$; and $\hat{\mathbf{A}}_{E_t} \in \mathbb{R}^{H \times W \times P \times B}$ is the predicted acoustic parameter map for emitter location $\mathbf{E}_t$, with $P$ acoustic parameters computed over $B$ frequency bands.

\subsection{Feature and labels extraction}
\label{sec:method:features}

\textbf{Floormaps and Reference RIRs} In this work, we consider floormaps as a lightweight source of geometric information of the scene, which is consumed by our model.  Floormaps can be obtained easily for real-world rooms from building construction plans or site layouts, or simple manual measurements.  Other methods to extract floormaps without knowledge of the complete 3D geometry \cite{mura2021walk2map, gueze2023floor, majumder2023chat2map} have also been well-studied.  However, since we use synthetic scene data in this work, we use the 3D meshes of the scenes to generate the floormaps. To extract the floormaps we take the full 3D mesh of the scene, and create a 2D map by slicing at a specified height. Our goal is to capture scene boundaries and internal subdivisions, but avoid details that have little impact on the late reverberation, like furniture. In practice, we use a fixed slice height of about 0.5 meter below the ceiling of the scene. The selected slice is then digitized into a binary 2D map of size $128\times128$. To provide acoustic context in addition to the geometric context, we also provide a reference RIR, from an arbitrary source and receiver position, as input to the model.  We encode these positions by marking their locations on another 2D binary map (via a transposed conv that places a Gaussian kernel at the location rather than a single active pixel), concatenated as an additional channel to the floormap. Finally, we compute the magnitude spectrogram of this reference IR and concatenate it as the final channel. We truncate the RIRs to 1 second, at 24 kHz and use 128 Mel bins with a hop size of 188 samples so that the spectrogram also becomes a 128 $\times$ 128 matrix. Thus, all the input features represent a single, multi-channel image.

\begin{figure}[htb]
      \centering
      \includegraphics[trim={0cm 0cm 0cm 0cm},clip, width=0.98\columnwidth]{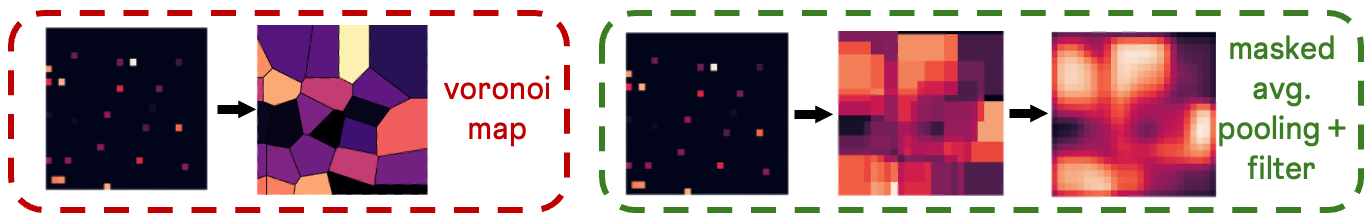}
    \captionsetup{width=0.98\linewidth}
    \caption{(left) Voronoi map of parameter values at sparse receiver locations, which creates fragmented maps with hard transitions. (right) Our acoustic heatmap processing of the same input, using a low passed masked average pooling operation. }
    \label{fig:masked_avg_pooling}
\end{figure}

\textbf{Acoustic Heatmaps}
The acoustic heatmaps represent the spatially-distributed acoustic parameters in the scene, and serve as labels for supervised learning. These parameters are computed from RIRs captured at a discrete set of receivers. To map the parameters from these sparse locations into a continuous and smooth heatmap, we use masked average pooling, where only the pixels that are active contribute to the pooling operation. Finally, to ensure a smooth response, we apply a 2D low pass filter via convolution with a Gaussian kernel (of size 9x9 and standard deviation of 1.). An example of the masked average pooling operation is shown in \cref{fig:masked_avg_pooling}. Unlike a Voronoi map, this operation creates continuous maps without hard transitions. This process is repeated for each desired acoustic parameter and frequency band, and stacked along the channel dimension. 

The acoustic parameters we use are well established. We focus on two main categories: 1) energy decay rates, that measure the time it takes for the energy of the RIR to decay to a specific level; and 2) early-to-late energy ratios, that measure the ratio of energy between the early reflections and the late reverberation at a predefined transition point. We compute the parameters following the methods as defined in the standard \cite{iso3382}. For our experiments, we focus on \paramEDT, \paramT{30} for decay rates and \paramDRR, \paramC{50} for energy ratios, which can be considered statistically sufficient to describe human perception of late reverberation in indoor environments \cite{helmholz2022towards, klein2023just, neidhardt2022}. Nevertheless, our approach is flexible and could be applied to other acoustic parameters.

\subsection{Floormaps to Acoustic Heatmaps}

We approach this task as an image translation task. In computer vision, the image-to-image translation involves transforming an image from one domain to another while preserving its original content (e.g. translating photographs to hand drawn sketches) \cite{pix2pix2017, CycleGAN2017, PangLQC22, Wang0ZTKC18}. Here we learn a mapping from the floormap and reference IR features to acoustic heatmaps that show the spatial distribution of acoustic parameters in the scene. To simplify the task, we set the target emitter $\mathbf{E}_t = \mathbf{E}_r$.   \cref{fig:system} shows the overall proposed solution. The network consumes a 2D floormap and reference RIR from an arbitrary source-receiver position as input features, and outputs a stack of heatmaps.  Each heatmap corresponds to one acoustic parameter (of \paramC{50}, \paramDRR, \paramT{60}, \paramEDT) at one frequency band (of 125, 250, 500, 1k, 2k, 4k Hz). 






More formally, we construct and train a model via:
\begin{equation}
\underset{\Phi}{\arg\min}\sum_{d \in D}\mathcal{L}\big(\Phi (\mathbf{F}_D, \mathbf{E}_r, \mathbf{R}_r, \mathbf{h}_r, \mathbf{E}_t), \mathbf{A}_{E_t}\big)
\end{equation}
where the loss is the pixel-wise mean absolute error between the target acoustic heatmaps and the predicted maps $\hat{\mathcal{A}}_{E_t}$, computed across $P$ acoustic parameters and $B$ frequency bands, defined as:
\begin{equation}
    \label{eq:loss}
    \mathcal{L}:= \sum_{p \in P} \sum_{b \in B} \big | (\mathbf{A}_{E_t} - \hat{\mathbf{A}}_{E_t}) \odot M \big | , 
\end{equation}
where $M$ is a binary mask where pixels with valid values (i.e., within the scene, with valid acoustic data, etc) in the target heatmap $\mathcal{A}_{E_t}$ are set to 1, and 0 otherwise. The element-wise multiplication $(\odot)$ ensures that the loss is computed only for valid pixels. The model $\Phi$ is a typical U-Net neural network based on ResNet blocks. 

\section{Datasets}

We consider several public datasets that include both scene geometry and RIRs, including \cite{gotz2021, GWA, prawda_karolina_2022_6985104}. However, we require a dataset with a large number of unique scene geometries, with multiple sources per scene, a dense grid of receivers, and multi-channel responses. In addition, we require diverse scene geometries and acoustics. We have a particular interest in multi-room scenes that represent typical indoor apartments, which are known for complex late reverberation phenomena \cite{mckenzie2021, bilon2006}. Therefore, we conduct experiments using one existing state-of-the-art dataset, SoundSpaces \cite{henJSGAIRG20}, and construct a novel dataset, the Multi-room Apartments Simulation (MRAS) dataset.

\subsection{Soundspaces}
We first use the Soundspaces 1.0 dataset \cite{henJSGAIRG20} with the Replica \cite{Replica} scenes, to enable comparisons with previous work. Replica has 18 scenes in total, including 3 multi-room apartments, a studio apartment with 6 different furniture configurations, and other small, single-room, shoe-box scenes. They typically contain about 250 sources and receivers, for a total of $250^2 ~= 60,000$ unique RIRs per scene.  However, acoustic diversity is limited; e.g. most scenes have \paramT{30} of around 0.6 seconds, with little variance across and within scenes. 


\subsection{Multi-Room Apartments Simulation (MRAS)}

The Multi-Room Apartments Simulation (MRAS) dataset is a novel multi-modal dataset created specifically for the task of estimating spatially-distributed acoustic parameters in complex scenes. It includes a large collection of scene geometries, with dozens of unique source positions, and a dense grid of receivers. The scene geometries are generated algorithmically by connecting shoe-box rooms using two distinct patterns. A key contribution of this work is the release of the MRAS dataset for public use. This includes the 3D meshes and the raw RIRs, as well as the pre-processed floormaps, acoustic parameters, and acoustic maps used in the experiments. 

\textbf{Scene Generation:}
We generate a total of 1000 scenes: 100 unique geometries using a linear pattern, 100 unique geometries using a grid pattern; each geometry is given 5 different sets of materials. For a total of $100\times2\times5 = 1000$ distinct acoustic scenes. The line pattern is built by connecting shoe-box rooms along a common boundary creating coupled-room scenarios. The grid pattern subdivides a large rectangular area into multiple connected rooms. In both cases, individual shoe-box rooms may have varying heights. Materials are randomly assigned to the floor, ceiling, and walls of each room. The materials are uniformly sampled from a set of realistic materials (e.g. carpet, concrete), plus two additional materials: a highly absorptive material (absorption $>$ 0.9 for all bands), and a highly reflective material (absorption $<$ 0.1 for all bands). Lastly, the doorframes that connect the rooms in a scene have random width, from a minimum of 0.9 meters up to as wide the wall it is located in. This creates scenes that can have wide hallways instead of only rooms connected via small openings. Although the scenes are constructed algorithmically, the geometries offer high acoustical complexity.




\textbf{Acoustic simulation:}
The dataset has approximately 4 million RIRs, divided across 1000 scenes. For each scene, we uniformly sample 3 receiver positions per room to act as source positions, regardless of the room size. We use a dense grid of receivers with 0.3 m of spacing, at least 0.5 m away from any boundary. The RIRs are in 2nd-order ambisonics, using the same ray-tracing methods as in \cite{henJSGAIRG20}. 


\section{Experiments}
\label{sec:experiments}

\begin{table*}[tb]
\scriptsize
\centering
\sisetup{
detect-weight, 
mode=text, 
tight-spacing=true,
round-mode=places,
round-precision=2,
table-number-alignment=center
}
\caption{Results for the prediction of 4 omnidirectional acoustic parameters aggregated over 6 frequency bands, (mean and standard deviation). 
} 
\begin{tabular}{@{}l c c *{12}{S@{\( \pm \)}S[table-format=2.2]}}

 \specialrule{1.0pt}{0.1pt}{3pt}

Model & Dataset & Fold &  
 \multicolumn{2}{c}{\paramC{50} (dB) $\downarrow$} & 
 \multicolumn{2}{c}{\paramT{30} (\%) $\downarrow$} & 
 \multicolumn{2}{c}{\paramDRR$ $ (dB) $\downarrow$} & 
 \multicolumn{2}{c}{\paramEDT$ $ (\%) $\downarrow$} & 
 \multicolumn{2}{c}{loss $\downarrow$} & 
\multicolumn{2}{c}{\text{SSIM} $\uparrow$}    \\
\midrule

\textsc{avg rir} &  Replica               & 1,2,4     &	   11.480475425720215 & 5.624965190887451 &	 33.566150069236755 & 13.955822587013245 &	 7.301405429840088 & 5.628262042999268 &	 69.31424140930176 & 26.80833637714386 &	 0.7100822925567627 & 0.6075302958488464 &	 0.14226792752742767 & 0.08230351656675339  \\

\textsc{scene avg rir}   &   Replica     & 1,2,4     &     	 9.802249908447266 & 4.340880870819092 &	 27.02738642692566 & 10.183553397655487 &	 6.820080280303955 & 5.048664093017578 &	 69.39169764518738 & 26.71731114387512 &	 0.6571704745292664 & 0.6025508642196655 &	 0.13866227865219116 & 0.08868491649627686 			   \\

\textsc{input rir}   &  Replica  & 1,2,4    &     3.821729898452759 & 2.377225399017334 &	 16.907411813735962 & 7.747884839773178 &	 2.605978012084961 & 1.3725332021713257 &	 38.135671615600586 & 20.090627670288086 &	 0.21929261088371277 & 0.17730417847633362 &	 0.3036629557609558 & 0.12005095928907394 \\

\textsc{scene random map}   &   Replica  & 1,2,4  &    3.5123161270401835 & 1.4841172348914058 &	 8.103388537671959 & 5.600528454391013 &	 2.367355071209018 & 0.9369375115087302 &	 21.192089386367488 & 12.246767551265371 &	 0.15303989241706148 & 0.09716598151272621 &	 0.39640224425367315 & 0.09645793449739426      \\

\textsc{scene avg map}   &   Replica  & 1,2,4 &       2.592002130351053 & 0.9276502098331325 &	 \textbf{5.86} & \textbf{4.21} &	 1.7147564182149777 & 0.6281534250696655 &	 \textbf{15.25} & \textbf{14.99} &	 \textbf{0.10} & \textbf{0.07} &	\textbf{0.54} & \textbf{0.07} 			  \\
\midrule

Ours   &   Replica               & 1,2,4     &	   \textbf{ 1.73} & \textbf{0.85} &	 10.771257430315018 & 5.8534279465675354 &	 \textbf{1.37} & \textbf{0.48} &	 17.01 & 10.16 &	 \textbf{0.10} & \textbf{0.06} &	 0.5019334554672241 & 0.08198314905166626    \\

Ours+NoRir  &   Replica      & 1,2,4         &	   	 1.8181877136230469 & 0.9359753727912903 &	 11.394183337688446 & 6.579876691102982 &	 1.4100247621536255 & 0.45601367950439453 &	 18.27634871006012 & 10.786548256874084 &	 0.10474103689193726 & 0.05767171457409859 &	 0.4811265170574188 & 0.09656191617250443			   \\
Ours+FloormapNoise(10pix)   &   Replica   & 1,2,4    &	  	 1.8471128940582275 & 0.8456493020057678 &	 10.905821621417999 & 5.0337761640548706 &	 1.4366272687911987 & 0.4690527617931366 &	 17.909568548202515 & 09.851040691137314 &	 0.10996966063976288 & 0.0654134452342987 &	 0.47184252738952637 & 0.01034474465996027 \\

\specialrule{0.8pt}{0.3pt}{3pt}

\textsc{avg rir}    &  MRAS               & 1 &  3.861624002456665 & 1.7218176126480103 &	 59.22422409057617 & 52.87929773330688 &	 2.1137855052948 & 0.8339098691940308 &	 40.35000503063202 & 26.162299513816833 &	 0.2208721786737442 & 0.13769370317459106 &	 0.4563758671283722 & 0.0937429815530777 \\

\textsc{scene avg rir}   &   MRAS     & 1 &  3.4441938400268555 & 1.647182583808899 &	 23.415949940681458 & 19.079698622226715 &	 2.206918239593506 & 1.0329440832138062 &	 32.718899846076965 & 20.60442715883255 &	 0.17027193307876587 & 0.10138069093227386 &	 0.4736173748970032 & 0.08255184441804886 \\

\textsc{input rir}   &  MRAS  & 1    &  3.501810312271118 & 2.222729444503784 &	 25.90664327144623 & 19.23666000366211 &	 2.4243264198303223 & 1.2566440105438232 &	 43.11063587665558 & 36.144256591796875 &	 0.18838511407375336 & 0.13342826068401337 &	 0.45795637369155884 & 0.10124801099300385 \\

\textsc{scene random map}   &   MRAS  & 1 &   2.225198957643984 & 0.9766155024510297 &	 \textbf{11.50} & \textbf{8.62} &	 1.3960278893207683 & 0.5303315602781543 &	 20.417761442989243 & 13.535384201784237 &	 \textbf{0.10} & \textbf{0.06} &	 \textbf{0.65} & \textbf{0.09}  \\

\textsc{scene avg map}   &   MRAS  & 1 &   2.9321074582900657 & 1.769801402857567 &	 14.455204422281354 & 1.4169081543119655 &	 1.87847023760725 & 0.9417561195617735 &	 25.76409369766944 & 24.61624186989634 &	 0.13303271434700376 & 0.10440649880740478 &	 0.5383984979626151 & 0.15416334805173876  \\

\midrule
Ours      & MRAS & 1 &  \textbf{1.87} & \textbf{0.90} &	 16.593855619430542 & 9.953510016202927 &	 \textbf{1.33} & \textbf{0.44} &	 \textbf{19.34} & \textbf{11.61} &	 \textbf{0.10} & \textbf{0.06} &	 0.5831785202026367 & 0.07684126496315002  \\

Ours+NoRir  &   MRAS     & 1         &	   2.7442708015441895 & 1.5500324964523315 &	 31.560856103897095 & 20.68723440170288 &	 1.6665077209472656 & 0.7380780577659607 &	 30.08742928504944 & 19.465671479701996 &	 0.1610836535692215 & 0.1199338510632515 &	 0.5333163142204285 & 0.10483009368181229  	 		   \\

Ours+FloormapNoise(10pix)     &   MRAS   & 1     &	  2.0938968658447266 & 1.0954744815826416 &	 19.410523772239685 & 10.244102030992508 &	 1.5114965438842773 & 0.4901463985443115 &	 23.224109411239624 & 13.091576099395752 &	 0.12029588222503662 & 0.06674448400735855 &	 0.48537805676460266 & 0.01603584922850132 \\

\bottomrule
\end{tabular}
\label{table:one}
\vspace{-3mm} 
\end{table*}

\subsection{Baselines}

Predicting acoustic parameters for a full, unseen scene in a single inference step is a new task; previous learning approaches to acoustics estimation have focused on within scene interpolation \cite{LuoDT00G22, SuCS22} or require rich, multimodal inputs \cite{mesh2ir, Wan2024, he2024deep, MajumderCAG22}.  To contextualize our approach and provide a more fair comparison, we compare the performance of our model to multiple algorithmic baselines. The baselines are ordered in increasing order of oracle information available to compute the result. First, we have baselines based on sampling an RIR from the dataset and computing the acoustic parameters on this RIR. The main difference between them is how the RIR is sampled. For \textbf{Average RIR} \textsc{(avg rir)} we randomly sample 500 RIRs from the dataset, compute the samplewise mean (time domain average) to create an overall representative RIR, such that all scenes have the same value for all pixels. For \textbf{Average RIR, Same Scene} \textsc{(scene avg rir)}, we repeat the process, but the sampling is done per scene; therefore each scene has its own representative RIR. For \textbf{Input RIR} \textsc{(input rir)} we take the same RIR as the proposed model. This baseline can be considered a fair baseline, as it has access to the same acoustic information as the proposed model.

We also include baselines based on sampling acoustic maps directly. For \textbf{Random Acoustic Map, Same Scene} \textsc{(scene rand map)} we sample the map of a random source per scene. For \textbf{Average Acoustic Map, Same Scene} \textsc{(scene avg map)} we sample up to 100 sources per scene, and compute the pixelwise mean. This is the strongest baseline as it uses information from all sources and receivers.


\subsection{Performance Evaluation}

We report performance on each acoustic parameter and additional metrics. For \paramC{50} and \paramDRR$ $, we measure the absolute error (dB), and for \paramEDT$ $  and \paramT{30} the proportional error (\%). These  are computed pixel-wise, only for pixels where valid measurement data exists. However, because these metrics do not consider any spatial structure across the map, we also include the Structural Similarity Index (SSIM) \cite{DBLP:journals/tip/WangBSS04}. 

Finally, we consider just noticeable difference (JND) metrics of 1 dB for energy ratio-based parameters (\paramC{50} and \paramDRR), and $10\%$ for decay time metrics (\paramT{30} and \paramEDT) $ $ \cite{BRADLEY199999, Werner2014}. Although true JNDs are difficult to define \cite{klein2023just}, and depend on several factors (e.g. frequency band, estimulus type, sound level, directivity, scene acoustics), they provide a useful sanity check on our results.

\subsection{Experimental Setup}
We split the datasets into train/test partitions by scenes, where each scene is only available in either train or test. For Replica, we manually select the scenes for each split, to keep a balanced distribution of scales and geometries. We train the model with a batch size of 128, minimizing the L1 loss \cref{eq:loss}, using the ranger optimizer \cite{wright2021ranger21} with a learning rate of 1e-3, until the validation loss stops decreasing for 3 consecutive validation steps. All acoustic parameters are normalized such that 90\% of the values fall in the (-1, 1) range. 
As data augmentation, we use random centered rotations and translations of the floormaps, constrained such that the whole scene is always visible in the floormap.

\subsection{Results}

\begin{figure*}[b!]
      \centering
      \includegraphics[trim={0.1cm 0cm 0.1cm 0.1cm},clip, width=0.90\linewidth]{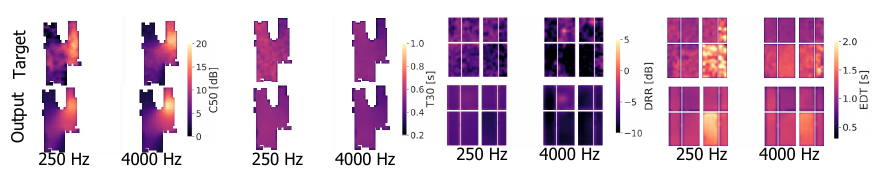}  
    \captionsetup{width=0.95\linewidth}
    \caption{Example of target heatmaps, and outputs of the proposed model for (2 left) Replica, and (2 right) multi-room scenes.}
    \label{fig:results_main}
\end{figure*}

\cref{table:one} shows the performance of the proposed model compared to the baselines on both datasets, for 4 acoustic parameters (\paramC{50}, \paramDRR, \paramT{30} and \paramEDT) at 6 frequency bands. First, we notice that the performance of the baselines mostly follows the amount of information available to them, where RIR-based baselines perform the worst. Secondly, our model outperforms all baselines, with some nuances. For parameters based on energy ratio (\paramC{50}, \paramDRR) the model achieves between 0.5 and 1 dB less mean error than the best baseline. However, for reverberation time metrics (\paramT{30}, \paramEDT) the model is slightly worse than the best baseline, but still significantly better than \textsc{input rir} (which has the same acoustic information as our model). This is because for most scenes, reverberation time does not change significantly with source position. Therefore, the average of multiple acoustic maps of the same scene approximates the reverberation time quite well. Removing the reference RIR (Ours+NoRIR) has little impact in Replica, but a large one in MRAS. This is due to the limited acoustic diversity in Replica. The model is robust to noisy floormaps (Ours+FloormapNoise), where adding random pixel displacement to the floormaps up to 10 pixels (3 meters), has only a moderate drop in performance.

Furthermore, the overall trends are consistent across both datasets with two main differences. Firstly, the performance for \paramC{50} and \paramDRR $\ $ on the RIR-based baselines is significantly better on the MRAS dataset as compared to Replica. This can be attributed to the complex multi-room geometries in MRAS, which include scenes with sparsely connected rooms. These geometries have cases where the direct path between the source and receiver is very long, leading to low-energy IRs dominated by reverberation. Secondly, for \paramT{30}, the baselines are higher-performing on Replica than on MRAS. This is because MRAS exhibits much higher acoustic variance, leading to a wider range in ground truth \paramT{30} values. Despite these differences, the proposed model performs consistently across both datasets. 


An example is shown in \cref{fig:results_main}. The ground truth \paramC{50} and \paramDRR $ $ show strong dependency on the proximity to the source, while \paramT{30} is much more uniform across the scene. In contrast, our model successfully captures patterns such as line of sight and source proximity. However, the output is smoother and lacks fine-grained spatial variations. 


\subsection{Spatially-Dependent Acoustic Parameters}
\label{sec:experiments:spatial}

The proposed method is flexible and can be adapted to different types of acoustic parameters. An interesting case is the use of spatially-dependent acoustic parameters, that overall can provide a more complete characterization of a scene by also considering the direction of sound \cite{campos2021, meyer2022, PRINN2025110220}. As an illustrative task, we predict a single parameter, \paramC{50}, at fixed directions from each location in the scene. To do this, we first take the 2nd-order ambisonic RIRs and beamform to 5 fixed orientations in the scene (azimuth only), about $72\deg$ apart from each other. A key modeling difference is the inclusion of the pose channel. Since we train the model using random rotations of the scenes, the model must know the canonical orientation of the scene, to determine the rotated orientation of the fixed directions used when beamforming the ground truth. To address this, we add an additional channel to the input features, represented as a single line, which rotates in accordance with the floormap. 


\cref{table:four} shows the results for the spatially dependent case. These show the same trends as in \cref{table:one} but with more limited performance overall (given the more difficult task), and with the heatmap-based baselines being noticeably better than the RIR-based baselines. Nevertheless, our model outputerms all baselines.
\begin{table}[htb]
\scriptsize
\centering
\sisetup{
reset-text-series = false, 
text-series-to-math = true, 
detect-weight, 
mode=text, 
tight-spacing=true,
round-mode=places,
round-precision=2,
table-number-alignment=center
}
\renewrobustcmd{\bfseries}{\fontseries{b}\selectfont}
\renewrobustcmd{\boldmath}{}
\newrobustcmd{\B}{\bfseries}
\addtolength{\tabcolsep}{-1.1pt}
\caption{Directional case. Prediction error of a single acoustic parameter, for 3 freq. bands, and 5 orientations (mean, std. dev.)} 
\begin{tabular}{@{}l c *{6}{S@{\( \pm \)}S[table-format=2.3]}}

 \specialrule{1.0pt}{0.1pt}{3pt}

Model  &  Dataset &
\multicolumn{2}{c}{\paramC{50} (dB) $\downarrow$} & 
\multicolumn{2}{c}{loss $\downarrow$} & 
\multicolumn{2}{c}{\text{SSIM} $\uparrow$} \\
  
\midrule

\textsc{avg rir}    &  Replica    &	 15.387764930725098 & 5.171406269073486 &  0.7548062801361084 & 0.26074841618537903 &	 0.3420429229736328 & 0.058866605162620544 \\

\textsc{scene avg rir}   & Replica    &	  10.835715293884277 & 3.0119268894195557 & 
 0.5276052951812744 & 0.15342362225055695 &	 0.3909345865249634 & 0.05480726435780525 \\

\textsc{input rir}   &   Replica    &	   4.866123199462891 & 2.392495632171631 &   0.24060359597206116 & 0.11687781661748886 &	 0.475972056388855 & 0.07496945559978485 \\

\textsc{scene random map}   &   Replica    &	 4.107126202843897 & 1.568736006152955 &  0.20984105590793764 & 0.07984479153870916 &	 0.46093671354936144 & 0.11223684277669867 \\

\textsc{scene avg map}   &   Replica     &   3.093550454519795 & 0.8919499907667507 & 	 0.14434725199975385 & 0.04235185285060591 &	 0.6283211619136597 & 0.07773097561587491 \\

\midrule

Ours   &   Replica     &    3.1216814517974854 & 1.1574078798294067 &   
   0.15937839448451996 & 0.05970659852027893 &	 0.5729939937591553 & 0.07481874525547028     \\


Ours+pose   &   Replica         &    \B 1.9359946250915527 & \B 0.9244021773338318 &  \B 0.09681583940982819 & \B 0.04751081019639969 &	 \B 0.6650294661521912 & \B 0.10213721543550491 \\


\bottomrule
\end{tabular}
\label{table:four}
\vspace{-3mm} 
\end{table}

\section{Conclusion}
\label{sec:conclusion}


In this paper we explore the task of estimating acoustic parameters across an entire unseen scene, for an arbitrary source, in a single inference step. We present a method to jointly estimate multiple spatially distributed acoustic parameters at multiple frequency bands, using limited geometric and acoustic context as input.  We validate our approach on the Replica dataset, as well as a novel, large scale dataset, MRAS, comprised of complex multi-room indoor scenes. 




\clearpage
\bibliographystyle{ieeetr}
\bibliography{references}







\end{document}